\documentclass{cernyrep} \usepackage{texnames} \usepackage[T1]{fontenc}
\usepackage{graphicx} \usepackage{epstopdf} \usepackage{xcolor}
\usepackage{nopageno}

\usepackage{hyperref}
\newcommand{\RM}[1]{\mathrm{#1}}
\begin{document} \title{Cyclotrons for
high-intensity beams}

\author{Mike Seidel}

\institute{Paul Scherrer Institute, Villigen, Switzerland}

\maketitle 

\begin{abstract} This paper reviews the important physical and
technological aspects of cyclotrons for the acceleration of
high-intensity beams. Special emphasis is given to the discussion
of beam loss mechanisms and extraction schemes.
\end{abstract}

\section{Introduction}

Cyclotrons have a long history in accelerator physics and are used
for a wide range of medical, industrial, and research applications
\cite{onishenko}. The first cyclotrons were designed and built by
Lawrence and Livingston \cite{lawrence} back in 1931. The
cyclotron represents a resonant-accelerator concept with several
properties that make it well suited for the acceleration of hadron
beams with high average intensity. In this paper, we concentrate
on aspects of the high-intensity operation of cyclotrons. The
electronic version of this document contains, in the references
section, clickable links to many publications related to this
theme.

\section{The classical cyclotron}

\begin{figure}
\centering\includegraphics[width=0.35\textwidth]{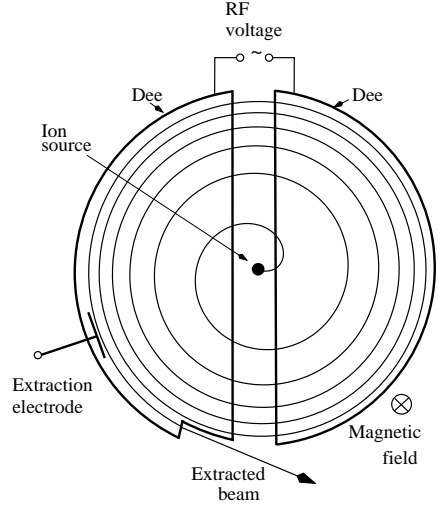}
\caption{Conceptual sketch of a classical cyclotron in plan view.
In the non-relativistic approximation, the turn separation scales
with the number of turns as $n_\mathrm{t}^{-1/2}$.}
\label{fig:spiral}
\end{figure}

Although the classical cyclotron has major limitations and is
practically outdated today, some fundamental relations are best
explained within this original concept. In the classical
cyclotron, an alternating high voltage at radio frequency (RF) is
applied to two D-shaped hollow electrodes, the \emph{dees}, for
the purpose of acceleration. Ions from a central ion source are
repeatedly accelerated from one dee to the other. The ions are
kept on a piecewise circular path by the application of a uniform,
vertically oriented magnetic field. On the last turn, the ions are
extracted by applying an electrostatic field using an electrode.
The concept is illustrated in Fig.~\ref{fig:spiral}.

The revolution frequency of the particle motion, called the
\emph{cyclotron frequency}, depends on the magnetic field $B_z$,
and the charge $q$ and the effective mass $\gamma m_0$ of the
particles:
\begin{eqnarray} f_\mathrm{c} = \frac{\omega_\mathrm{c}}{2\pi} & = & \frac{qB_z}{2\pi \gamma m_0}
\nonumber \\
& \approx & 15.2~ \RM{MHz}\cdot B \RM{(T)}
\RM{~(\,for\,\,protons\,)}.
\end{eqnarray}
The frequency of the accelerating voltage must be equal to the
cyclotron frequency or an integer multiple of it, i.e.,
$\omega_\RM{RF} = h \omega_\mathrm{c}$. The harmonic number $h$
equals the number of bunches that can be accelerated in one turn.
With increasing velocity, particles travel at larger radii, so
that $R \propto \beta$, and the revolution time remains constant
and in phase with the RF voltage. The bending strength is given by
$B_z R \propto p \propto \beta\gamma$. Thus, as long as $\gamma
\approx 1$, the condition of \emph{isochronicity} is fulfilled in
a homogeneous magnetic field. However, for relativistic particles,
the magnitude of the B-field has to be raised in proportion to
$\gamma$ at increasing radii, in order to keep the revolution time
constant throughout the acceleration process.  In summary, the
condition of isochonicity in a cyclotron requires the following
scaling of the orbit radius and bending field:
\begin{equation}
R \propto \beta,~~~B_z \propto \gamma \, .
\label{isochron}
\end{equation}

This original cyclotron concept exhibits some essential properties
that allow the high-intensity application of cyclotrons, which is
the focus of this article. The acceleration process takes place
continuously, and neither the RF frequency nor the magnetic
bending field has to be cycled. The separation of subsequent turns
allows continuous extraction of the beam from the cyclotron.
Consequently, the production of a continuous-wave (CW) beam is a
natural feature of cyclotrons. The so-called $K$-value is a
commonly used parameter for the characterization of the magnetic
energy reach of a cyclotron design. This equals the maximum
attainable energy for protons in the non-relativistic
approximation. The $K$-value is proportional to the maximum
squared bending strength, i.e., $K \propto (B\rho)^2$, and can be
used to rescale the achievable kinetic energy per nucleon for
varying charge-to-mass ratio:
\begin{equation} \frac{E_\mathrm{k}}{A} = K \left( \frac{Q}{A} \right)^2.
\end{equation}

The radial variation of the bending field in a classical cyclotron
generates focusing forces. At a radius $R$, the slope of the
bending field is described by the field index $k$, where
\begin{equation} k = \frac{R}{B_z} \frac{\mathrm{d}B_z(R)}{\mathrm{d}R}.
\label{index} \end{equation} Using Eq.~(\ref{isochron}), the
scaling of the field index under isochronous conditions can be
evaluated as follows:
\begin{eqnarray}
\frac{R}{B} \frac{\mathrm{d}B}{\mathrm{d}R} & = &
\frac{\beta}{\gamma} \frac{\mathrm{d}\gamma}{\mathrm{d}\beta}
\nonumber \\
& = & \gamma^2 -1.
\label{index1}
\end{eqnarray}
The radial equation of motion of a single particle can be written
as
\begin{equation} m\ddot{r} = m r \dot{\varphi}^2-q r\dot{\varphi} B_z .
\end{equation}

We now consider small deviations around the central orbit $R$,
namely $r=R+x,x \ll R$:
\begin{eqnarray} \ddot{x} + \frac{q}{m} v B_z(R+x) - \frac{v^2}{R+x} & = &
0, \nonumber \\ \ddot{x} + \frac{q}{m} v \left( B_z(R) +
\frac{\mathrm{d}B_z}{\mathrm{d}R} x\right) -\frac{v^2}{R} \left(
1-\frac{x}{R} \right) & = & 0, \nonumber \\ \ddot{x} +
\omega_\mathrm{c}^2 (1+k) x & = & 0. \label{radmotion}
\end{eqnarray}
In this derivation, we have used the relations $\omega_\mathrm{c}
= qB_z/m \approx v/R$ and $r\dot{\varphi} \approx v$. Thus, in the
linear approximation, the horizontal `betatron motion' is a
harmonic oscillation around the central beam orbit, $x(t) =
x_\RM{max} \cos(\nu_r \omega_\mathrm{c} t)$. The parameter $\nu_r$
is called the betatron tune. From Eq.~(\ref{radmotion}), we see
that the radial betatron frequency in a classical cyclotron is
given by
\begin{eqnarray}
\nu_r & = & \sqrt{1+k} \nonumber \\
& \approx & \gamma.
\label{nu_r}
\end{eqnarray}
In the above, Eq.~(\ref{index1}) has been used to derive the
relation for $\gamma$. A similar calculation can be done for the
vertical plane, using Maxwell's equation $\RM{rot} \, \vec{B}=0$.
This yields the following for the vertical betatron frequency:
\begin{equation}
\nu_z = \sqrt{-k}.
\label{nu_z}
\end{equation}

Beta functions can be defined in the sense of the Courant--Snyder
theory \cite{courant} for a cyclotron. In the radial plane, the
average beta function can be estimated via
\begin{equation}
\beta_r \approx \frac{R}{\nu_r} \approx \frac{R}{\gamma}.
\label{beta}
\end{equation}
A radial dispersion function can be defined as well:
\begin{equation}
\Delta R = D_r \frac{\Delta p}{p},~~~D_r \approx
\frac{R}{\gamma^2}. \label{dispersion}
\end{equation}
The derivation of this relation is done in a similar way to the
calculation of the radial step width in Eq.~(\ref{step_1}) in the
next section. The two relations above can be used to establish
rough matching conditions for beam injection into cyclotrons. As
is obvious from Eq.~(\ref{nu_z}), vertical focusing can be
obtained only if the bending field decreases towards larger radii.
However, a negative slope of the field would be inconsistent with
the isochronicity condition stated above, which requires the field
to increase in proportion to $\gamma$. Thus the classical
cyclotron is limited to relatively low energies. As we will see in
the next section, vertical focusing can in fact be achieved by an
azimuthal variation of the bending field.

\section{AVF and separated-sector cyclotrons} \label{sec:avf}

\begin{figure}[h!]
\centering\includegraphics[width=0.65\textwidth]{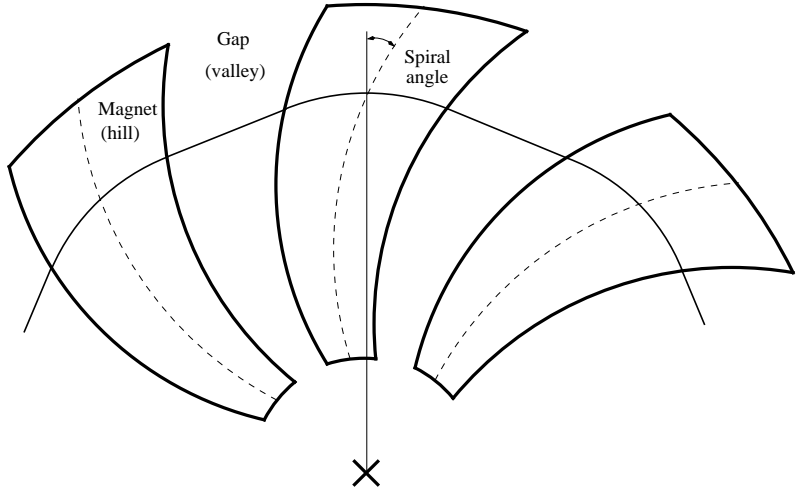}
\caption{Spiral magnet sectors and definition of the average
spiral angle} \label{fig:spiral_angle} \end{figure}

One way to overcome the problem of insufficient vertical focusing
is the introduction of azimuthally varying fields, as is done in
the Thomas, or AVF, cyclotron. The principle was proposed in 1938
by L.H.~Thomas~\cite{thomas}, but it took several decades before a
cyclotron based on this principle was actually built (in Delft in
1958). The variation of the vertical bending field along the
flight path leads to transverse forces on the particles that can
be utilized to provide suitable focusing characteristics in both
of the transverse planes. In a Thomas cyclotron, the average field
strength can be increased as a function of radius without losing
the vertical stability. As a result, this concept allows higher
energies to be achieved, for example 1~GeV for protons. In
fixed-focus alternating-gradient (FFAG) rings, the same type of
edge focusing plays a role as well; however, in most FFAG designs,
the focusing effect of the alternating gradients is dominant. The
required field variation in the cyclotron can be achieved by
special shaping of the poles of a compact single magnet. The
focusing can be increased by the introduction of spiral sector
shapes, and sector boundaries that have tilt angles with respect
to the beam orbit. With such magnet configurations, the squared
vertical betatron frequency is approximately
\begin{equation}
\nu_z^2 \approx -k + F^2 ( 1 + 2 \tan^2 \delta), ~~F^2 =
\frac{\overline{B_z^2}-\overline{B_z}^2}{\overline{B_z}^2}.
\label{flutter}
\end{equation}
The so-called flutter factor $F$ equals the relative root mean
square (r.m.s.) variation of the bending field around the
circumference of the cyclotron. The spiral angle $\delta$ is
defined as shown in Fig.~\ref{fig:spiral_angle}.

The next and most recent step in the history of cyclotron
development was the introduction of separated-sector cyclotrons.
Such cyclotrons have a modular structure consisting of several
sector-shaped dipole magnets and RF resonators for acceleration.
The modular concept makes it possible to construct larger
cyclotrons that can accommodate the bending radii of ions at
higher energies.

For completeness, the synchrocyclotron, which represents another
way to overcome the relativistic limit, should also be mentioned
here. In the synchrocyclotron, the RF frequency is varied
according to the variation in the speed of the accelerated
particles. This means that pulses, i.e., trains of bunches of
ions, are accelerated, which results in a drastic reduction in the
average beam current that can be achieved. Historically, the
synchrocyclotron was an important step in pushing the energy
frontier, but this concept was superseded by the invention of the
synchrotron. The sector-focused cyclotron is limited in energy,
but it has retained its attractiveness for high-intensity
applications owing to its advantage of CW operation. A
comprehensive overview article of cyclotron concepts can be found
in \cite{onishenko}.

For clean extraction with an extraction septum, the distance
between the turns at the extraction radius must be maximized. It
is therefore essential in the design of a high-intensity cyclotron
to consider the transverse separation of the beam. To calculate
the step width per turn, we start from the formula for the
magnetic rigidity,
\begin{equation} BR = \frac{p}{e} = \sqrt{\gamma^2-1} ~\frac{m_0c}{e}.
\end{equation}
By computing the total logarithmic differential, we obtain a
relation between the changes in the radius, magnetic field, and
energy of the particle beam:
\begin{equation} \frac{\mathrm{d}B}{B} +\frac{\mathrm{d}R}{R} = \frac{\gamma~\mathrm{d}\gamma}{\gamma^2-1}.
\end{equation}
Using the field index $k$ given in Eq.~(\ref{index}), we obtain
\begin{equation} 1+k = \frac{\gamma R}{\gamma^2 -1} \frac{\mathrm{d}\gamma}{\mathrm{d}R}.
\nonumber \end{equation}
Noting that the change in the relativistic quantity $\gamma$ per
turn is $\mathrm{d}\gamma/\mathrm{d}n_\mathrm{t} =
U_\mathrm{t}/(m_0c^2)$, where $U_\mathrm{t}$ denotes the energy
gain per turn, we finally obtain the step width in the radius,
\begin{eqnarray}
\frac{\mathrm{d}R}{\mathrm{d}n_\mathrm{t}} & = & \frac{\mathrm{d}\gamma}{\mathrm{d}n_\mathrm{t}} \frac{\mathrm{d}R}{\mathrm{d}\gamma} \nonumber \\
& = & \frac{U_\mathrm{t}}{m_0c^2} \frac{\gamma R}{(\gamma^2-1)(1+k)} \label{step_1} \\
& = & \frac{U_\mathrm{t}}{m_0c^2} \frac{\gamma
R}{(\gamma^2-1)\nu_r^2}. \label{step_2}
\end{eqnarray}

\begin{figure}[h!]
\centering\includegraphics[width=0.65\textwidth]{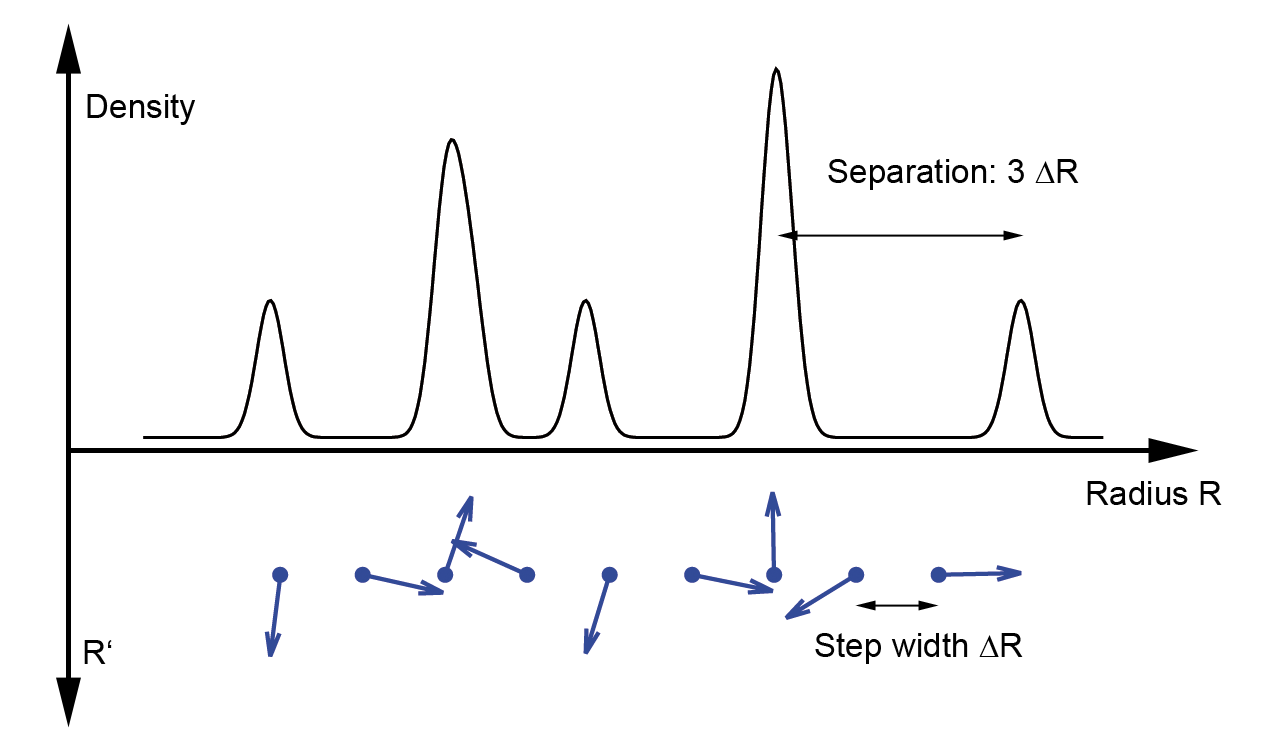}
\caption{Betatron oscillations of the centre of the beam around a
closed orbit can be utilized to maximize the beam separation at
extraction. The upper plot shows the calculated beam density,
which is a superposition of Gaussian profiles. In the lower half,
the clockwise-rotating phase space vector of the centroid of the
beam is shown for each turn.} \label{fig:steps} \end{figure}

In the outer region of the cyclotron, near the extraction radius,
it is possible to violate the condition of isochronicity for a few
turns. By reducing the slope of the field strength, which is
related to the radial tune, it is possible to increase the turn
separation locally. In the fringe field region of the magnets, the
field decreases naturally. By going from Eq.~(\ref{step_1}) to Eq.~(\ref{step_2}) using Eq.~(\ref{nu_r}), we can show the relation
between the step width and the radial tune. If the condition of
isochronicity remains valid, the dependence on the field index and
the radial tune can be eliminated, and the step width is given by
\begin{equation}
\frac{\mathrm{d}R}{\mathrm{d}n_\mathrm{t}}  =
\frac{U_\mathrm{t}}{m_0c^2} \frac{R}{(\gamma^2-1)\gamma}.
\label{step_full}
\end{equation}
In this form, the equation shows the strong dependence of the step
width on the beam energy. Above 1~GeV, it becomes very difficult
to achieve clean extraction with an extraction septum. An
effective way to increase the turn separation at the extraction
element is the introduction of orbit oscillations by deliberately
injecting the beam slightly off centre. When the phase and
amplitude of the orbit oscillation are chosen appropriately, and
also the behaviour of the radial tune is controlled in a suitable
way, the beam separation can be increased by a factor of three.
According to Eq.~(\ref{step_1}), this gain is equivalent to a
cyclotron three times larger and is thus significant. Figure~\ref{fig:steps} illustrates how this scheme is used in the PSI
Ring cyclotron. In~\cite{bi}, the beam profile in the outer turns
was computed numerically for realistic conditions, and the results
are in good agreement with measurements.

In summary, the clean extraction of the beam is of the utmost
importance for high-intensity cyclotron operation. The turn
separation at the extraction element can be maximized by the
following measures.
\begin{itemize}
\item The extraction radius should be large, i.e., the overall
dimensions of the cyclotron should not be chosen to be too small.
\item The energy gain per turn should be maximized by installing a
sufficient number of resonators with high performance. \item At
relativistic energies, the turn separation diminishes quickly, and
thus the final energy should be kept below approximately 1~GeV.
\item In the extraction region, the turn separation can be
increased by lowering the slope of the field index and by
utilizing orbit oscillations resulting from controlled off-centre
injection.
\end{itemize}

An alternative to the extraction method described here is
extraction via charge exchange. More details of this method are
given in Section \ref{sec:extraction}.

\section{Design aspects of separated-sector cyclotrons}

Modern cyclotrons that are able to reach higher $K$-values,
particularly those designed for high intensity, are typically
realized as separated-sector cyclotrons. They employ a modular
concept involving a combination of sector-shaped magnets, RF
resonators, and empty sector gaps to form a closed circular
accelerator. The modular concept simplifies the construction of
cyclotrons with diameters significantly larger than those achieved
with the classical single-magnet concept. The large orbit radius
at maximum energy permits extraction with extremely low losses.
The modularity also has significant advantages concerning the
serviceability of the accelerator, especially in view of the need
to handle activated components.

During the course of acceleration, the revolution time is kept
constant, which leads to a significant variation in the average
orbit radius. The lateral width of the elements in the ring is
large in comparison with, for example, the elements of a
synchrotron that uses strong focusing. The mechanical design of
the vacuum chambers and sealed interconnections is thus
challenging. On the other hand, the large variation in the radius
makes it possible to separate the turns at the outer radius and to
realize an extraction scheme for CW operation with very low
losses. The close orbit spacing in FFAG rings, for example, makes
continuous extraction difficult. The extraction loss is the
limiting effect for high-intensity operation of cyclotrons.

The wide vacuum chambers (2.5~m for the PSI Ring cyclotron)
require special sealing techniques. In a cyclotron, as in a
single-pass accelerator, vacuum levels of $10^{-6}$~mbar are
sufficient for the acceleration of protons. So-called inflatable
seals are manufactured from thin steel sheets with two sealing
surfaces per side and an intermittently evacuated volume between.
To simplify installation, these seals are positioned on radial
rails between two elements. Inflation with pressurized air seals
the surfaces. This screwless scheme can tolerate small positioning
errors and has the advantage of short mounting times.

\begin{figure}
\centering\includegraphics[width=0.95\textwidth]{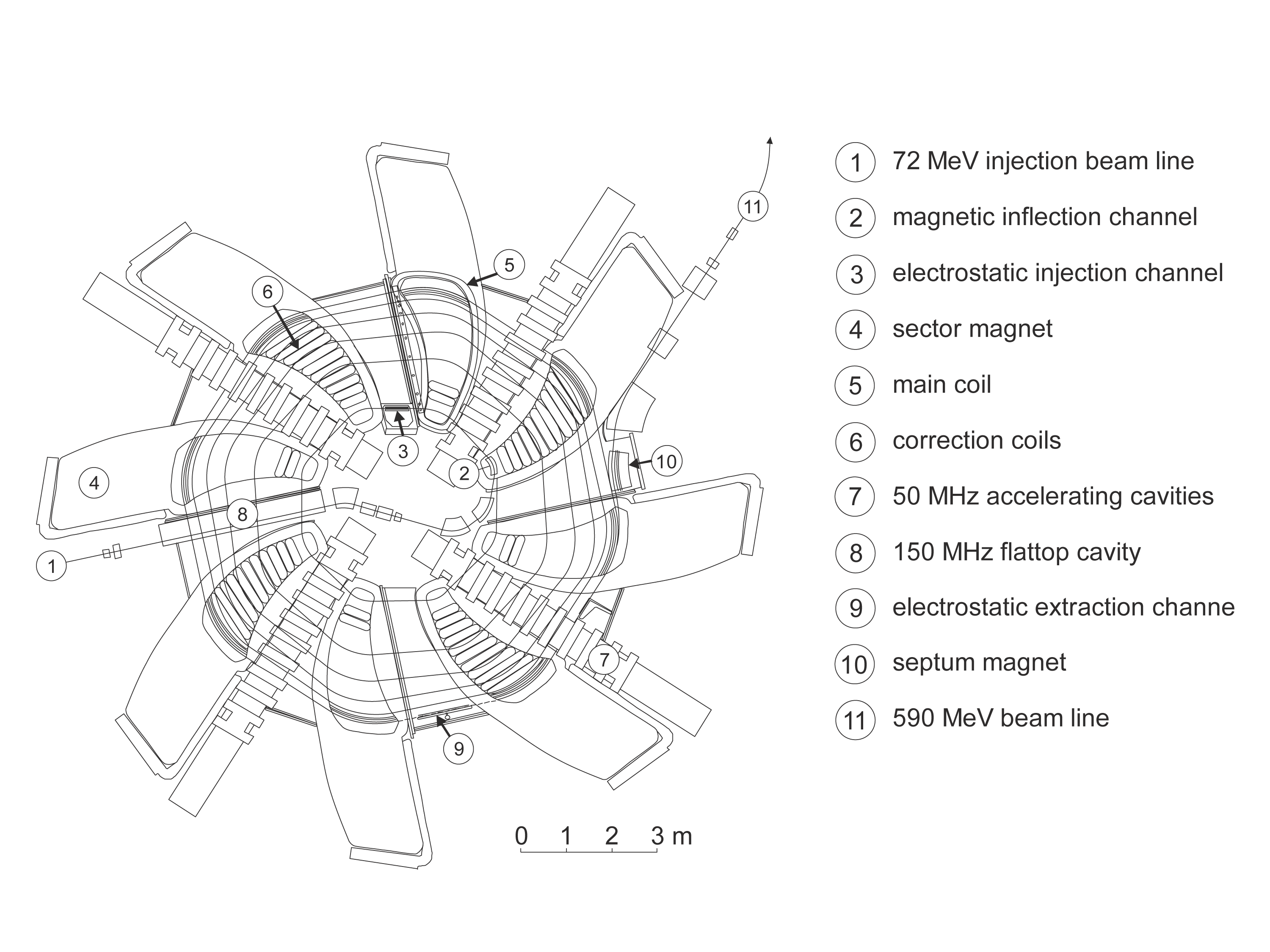}
\caption{Top view of the PSI Ring cyclotron. This separated-sector
cyclotron contains eight sector magnets, four accelerating
resonators (50~MHz), and one flat-top resonator (150~MHz).}
\label{fig:ring}
\end{figure}

The concept of the separated-sector cyclotron requires external
injection of a beam of good quality. Both injection and extraction
are often performed using an electrostatic deflection channel. In
both cases the beam is deflected at a certain radius, while the
neighbouring turns must not be affected. This is achieved by
placing a thin electrode between the two turns. Particles in the
beam tails that hit this electrode are scattered, and these
generate losses and activation. A magnetic element would need much
more material to be placed between the turns.  A simplified view
of the PSI Ring cyclotron is given in Fig.~\ref{fig:ring}.

Some parameters of large cyclotrons operating today are listed in
Table~\ref{tab:cyclotrons}. The TRIUMF cyclotron~\cite{dutto}
accelerates H$^-$ ions and allows the extraction radius to be
varied to adjust the final beam energy. The RIKEN Ring
cyclotron~\cite{kase} is not a high-intensity machine, but it
allows a broad variety of ions to be accelerated. The special
feature of the RIKEN cyclotron is the superconducting sector
magnets, which deliver a very high bending strength, reflected by
the corresponding $K$-value. The PSI Ring cyclotron was proposed
in the 1960s by Willax \cite{willax}. It is specialized for
high-intensity operation at the expense of reduced flexibility.

\begin{table} 
\caption{Selected parameters of large sector cyclotrons. The
TRIUMF cyclotron uses a single magnet with sector poles. The
maximum beam power of the RIKEN cyclotron was achieved in 2011,
and may change with other ion species or operation modes.}
\centering \begin{tabular}{ccccccccc}
\hline\hline
\rule{0mm}{6mm} Cyclotron & $K$ & $N_\RM{mag}$ & Harmonic & $R_\RM{inj}$ & $R_\RM{extr}$ & Extraction & Overall & $P_\RM{max}$ \\
\rule[-3mm]{0mm}{6mm} & (MeV) & & number & (m) & (m) & method & transmission & (W) \\
\hline
\rule{0mm}{6mm} TRIUMF & 520 & 6 & 5 & 0.25 & 3.8--7.9 & H$^-$ stripping & 0.70 & 110 \\
& & (sectors) & & & & channel & & \\
PSI Ring & 592 & 8 & 6 & 2.1 & 4.5 & Electrostatic & 0.9998 & 1400 \\
& & & & & & channel & & \\
RIKEN & 2600 & 6 & 6 & 3.6 & 5.4 & Electrostatic & 0.63 & 6.2 \\
\rule[-3mm]{0mm}{6mm} Ring & & & & & & channel & &
$\left( ^{18} \RM{O} \right)$ \\
\hline\hline
\end{tabular}

\label{tab:cyclotrons}
\end{table}

\section{Space charge effects in cyclotrons}

In high-intensity cyclotrons, space charge effects are of major
importance in determining the maximum attainable intensity. In
principle, the CW operation of cyclotrons results in low bunch
charges, leading to moderate space charge effects in comparison
with pulsed-accelerator concepts. On the other hand, the focusing
forces are rather weak. In the transverse planes, space charge
forces cause shifts in the focusing frequencies, and for large
tune shifts this results in resonant losses. Strong defocusing
space charge forces may even exceed the focusing forces generated
by the cyclotron magnets. Longitudinal space charge forces lead to
an increased energy spread, which is transformed into transverse
beam tails. In cyclotrons with a deflecting extraction element,
these beam tails are scattered at this element and limit the
intensity. Analytical prediction of the beam dynamics under the
influence of space charge forces is difficult, since the beams in
different turns overlap, and therefore forces from neighbouring
bunches cannot be neglected in general. With particle-tracking
codes, the self-induced fields of a bunch under consideration and
neighbouring bunches can be included in detailed predictions of
the beam dynamics \cite{jianjun}. However, in order to gain some
insight into the fundamental dynamics, it is sufficient to
consider a simplified model of uniformly charged beam sectors
resulting from completely overlapping turns.

For very short transversely separated bunches, the strong
repelling space charge force results in a rapid motion of the
particles around the centre of the bunch on cycloidal paths. In
this specific regime, a compact, stable circular bunch shape is
developed, despite the presence of a repelling central force
within the bunch. In this case, complete coupling between the
longitudinal and radial degrees of freedom is observed. Without
going into too much detail, we will briefly summarize the
transverse and longitudinal effects here, as well as the formation
of a round beam for short bunches.

\subsection{Transverse space charge forces}

In the case of a cyclotron with overlapping turns, a current sheet
model, assuming flat rotating sectors of charge, can be applied.
The vertical force on a test particle at a distance $y$ from the
beam centre is
\begin{equation}
F_y = \frac{n_v e^2}{\epsilon_0 \gamma^2} \cdot y. \label{vforce}
\end{equation}
Here, $n_v$ is the particle density in the centre of the bunch,
given by
\begin{equation}
n_v = \frac{N}{(2\pi)^{3/2} \sigma_y D_\mathrm{f} R \, \Delta R},
\label{vol_dens} \end{equation}
where $D_\mathrm{f}$ is the fraction of the circumference covered
by the beam, i.e., the ratio of the average current to the peak
current; $\sigma_y$ is the vertical r.m.s. beam size; $\Delta R$
is the step width between the turns, which was discussed in
Section \ref{sec:avf}; and $N$ is the number of particles per
turn, contained in $h$ bunches. Assuming a Gaussian longitudinal
distribution with an r.m.s. bunch length $\sigma_z$, we have
\begin{equation}
D_\mathrm{f} = \frac{h \sigma_z}{(2\pi)^{1/2} R} . \nonumber
\end{equation}
The focusing force generated by the magnet structure can be
expressed as
\begin{equation}
F_y = - \gamma m_0 \omega_\mathrm{c}^2 \nu_{y0}^2 \cdot y.
\label{force_focus} \end{equation}
Thus the resulting vertical equation of motion for a test particle
can be written as
\begin{equation}
\ddot{y} + \left( \omega_\mathrm{c}^2 \nu_{y0}^2 - \frac{n
e^2}{\epsilon_0 m_0 \gamma^3} \right) y = 0. \label{motion_spc}
\end{equation}

Obviously, an intensity limit is reached when the focusing term in
the brackets vanishes. This condition was used by Blosser
\cite{handbook} to formulate a space charge limit for cyclotrons.
In practice, an operating limit is reached somewhat earlier for a
tune shift of approximately 0.4. The effective vertical tune can
be deduced from Eq.~(\ref{motion_spc}) as follows:
\begin{eqnarray}
\nu_y & = & \left( \nu_{y0} - \frac{4\pi c^2 r_\mathrm{p} n_\mathrm{v}}{\omega_\mathrm{c}^2 \gamma^3} \right)^{1/2} \nonumber \\
& \approx & \nu_{y0} - \frac{2\pi c^2 r_\mathrm{p}
n_\mathrm{v}}{\omega_\mathrm{c}^2 \gamma^3 \nu_{y0}}.
\label{tshift_1}
\end{eqnarray}
Here, the classical proton radius $r_\mathrm{p} = 1.5\times
10^{-18}$~m has been introduced. After some algebra, and by
replacing the step width $\Delta R$ using Eq.~(\ref{step_full}),
we finally obtain the following for the tune shift:
\begin{equation}
\Delta \nu_y = - \sqrt{2\pi} \frac{r_\mathrm{p} R}{e\beta c
\nu_{y0} \sigma_z} \frac{m_0 c^2}{U_\mathrm{t}} ~I_\RM{avg}.
\label{tshift_2}
\end{equation}
For typical high-intensity cyclotrons, this formula predicts a
space charge limit at currents of several tens of milliamps.

\subsection{Longitudinal space charge forces}

Beam losses caused by longitudinal effects are important even at
the milliamp level. The effect of longitudinal space charge forces
can also be estimated in a sector model. Because of the shielding
effect of the vacuum chamber, only charges within a radius
approximately equal to the chamber height contribute to the force.
Joho~\cite{joho} estimated a quadratic dependence of the
accumulated energy spread on the turn number $n_\mathrm{t}$:
\begin{eqnarray}
\Delta E_\RM{sc} & = & \frac{16}{3} \frac{e g_{1\mathrm{c}} Z_0}{\beta_\RM{max}}~\frac{I_\RM{avg}}{D_\mathrm{f}} n_\mathrm{t}^2 \nonumber \\
& \approx & 2800 (\Omega) \frac{e I_\RM{avg}
n_\mathrm{t}^2}{D_\mathrm{f} \beta_\RM{max}}. \label{lspc}
\end{eqnarray}
Here, $Z_0 = 377~\Omega$ is the impedance of free space and
$g_{1\mathrm{c}} \approx 1.4$ is a form factor. The calculation
assumes non-relativistic conditions. In Ref.~\cite{pozdeyev}, the
results of numerical simulations were compared with this simple
analytical calculation and the agreement was satisfactory.

The energy spread generated by longitudinal space charge forces is
transformed into transverse beam tails. Losses occur at the
electrode of the extraction element owing to residual beam density
between the orbits of the last two turns. The separation between
these turns is proportional to $n_\mathrm{t}^{-1}$. Consequently,
under the constraint of constant losses, the maximum attainable
beam current scales in inverse proportion to the third power of
the turn number. Over the history of the PSI cyclotron
accelerator, the beam current has been increased by a large
factor. This has been achieved mainly by applying higher gap
voltages in the resonators, thus reducing the number of turns.
More powerful RF amplifiers and new resonators have been
installed. In fact, the maximum beam current scaled according to
the above third-power law~\cite{msei}.

\subsection{Circular-bunch regime}

\begin{figure}
\centering\includegraphics[width=0.95\textwidth]{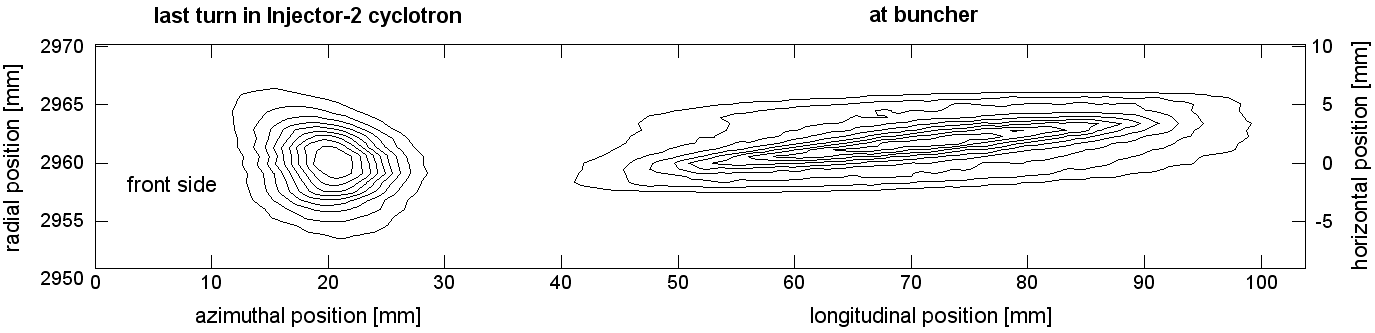}
\caption{Particle density for a current of 2.2~mA measured at the
exit of the PSI Injector~II cyclotron (72~MeV) and after a drift
length of $\approx20$~m. The profile was measured by detecting
protons scattered from a vertically oriented wire probe placed at
varying horizontal positions. The distribution of such events was
recorded as a function of time \cite{doelling}.}
\label{fig:timestructure}
\end{figure}

For very short bunches, a self-focusing effect can be observed in
a cyclotron, which leads to the formation of a circular stable
bunch shape in the radial--longitudinal plane. Owing to the
combination of strong space charge forces and the bending dipole
field, the particles start to rotate rapidly around the bunch
centre in cycloidal paths. Analytical descriptions of this effect
were given by Chasman and Baltz~\cite{chasman} for the case of a
potential due to a point-like central charge and by Bertrand and
Ricaud~\cite{bertrand} for the case of a potential due to a
constant charge density. The bunches in the individual turns must
be separated for most of the acceleration time in order to enter
the circular-shape regime. This behaviour obviously does not occur
in the sector model described earlier, involving overlapping
turns. A circular bunch shape is observed in practice in the
Injector~II cyclotron at the PSI. Figure~\ref{fig:timestructure}
shows the measured circular particle distribution at the exit of
Injector~II and, for comparison, the distribution measured roughly
20~m downstream. Over the relatively short drift length, the bunch
shows a significant longitudinal increase, whereas over more than
200~m travel distance in the cyclotron it stays in the compact
form shown. For a high-intensity cyclotron, it is of course
desirable to enter the circular-bunch regime, since the
longitudinal beam blow-up can be drastically reduced.

\section{Injection and extraction} \label{sec:extraction}

\begin{figure}
\centering\includegraphics[width=0.8\textwidth]{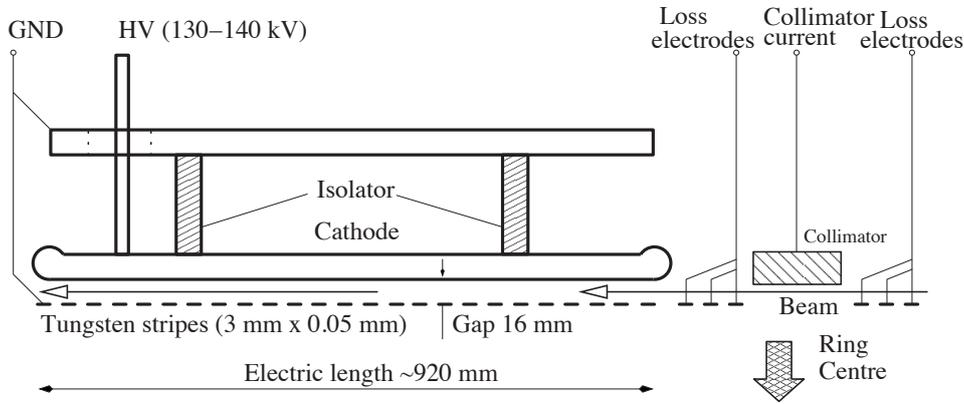}
\caption{Electrostatic extraction channel of the PSI Ring
cyclotron} \label{fig:eec} \end{figure}

Extraction of the beam at high energy is one of the critical
aspects of high-intensity cyclotrons. Two schemes are used for
extraction. In the more classical scheme, the electrode of an
electrostatic deflector is placed between the last and the second
last turn in the cyclotron. The beam receives a kick angle of the
order of 10~mrad, which is enough to separate the orbits to a
distance that allows the insertion of a septum magnet. Although a
thin electrode is typically used, some tail particles of the beam
hit the electrode. The scattered particles may end up in the
vacuum chamber in the extraction beam line. As described before,
for this extraction scheme to work it is important to generate a
large turn separation, resulting in a low beam density at the
location of the electrode. A schematic drawing of an electrostatic
deflector is shown in Fig.~\ref{fig:eec}. The deflection angle can
be calculated via the electrical rigidity,
\begin{eqnarray}
E\rho & = & \frac{\gamma + 1}{\gamma} \frac{E_\mathrm{k}}{q} \\
         & \approx & 2U_\RM{gap} ~~(\RM{for~}E_\mathrm{k}\ll E_0). \nonumber
\label{e_rigitity}
\end{eqnarray}
In the low-energy approximation, $U_\RM{gap}$ denotes the gap
voltage of the electrostatic deflector.

\begin{figure}
\centering\includegraphics[width=0.65\textwidth]{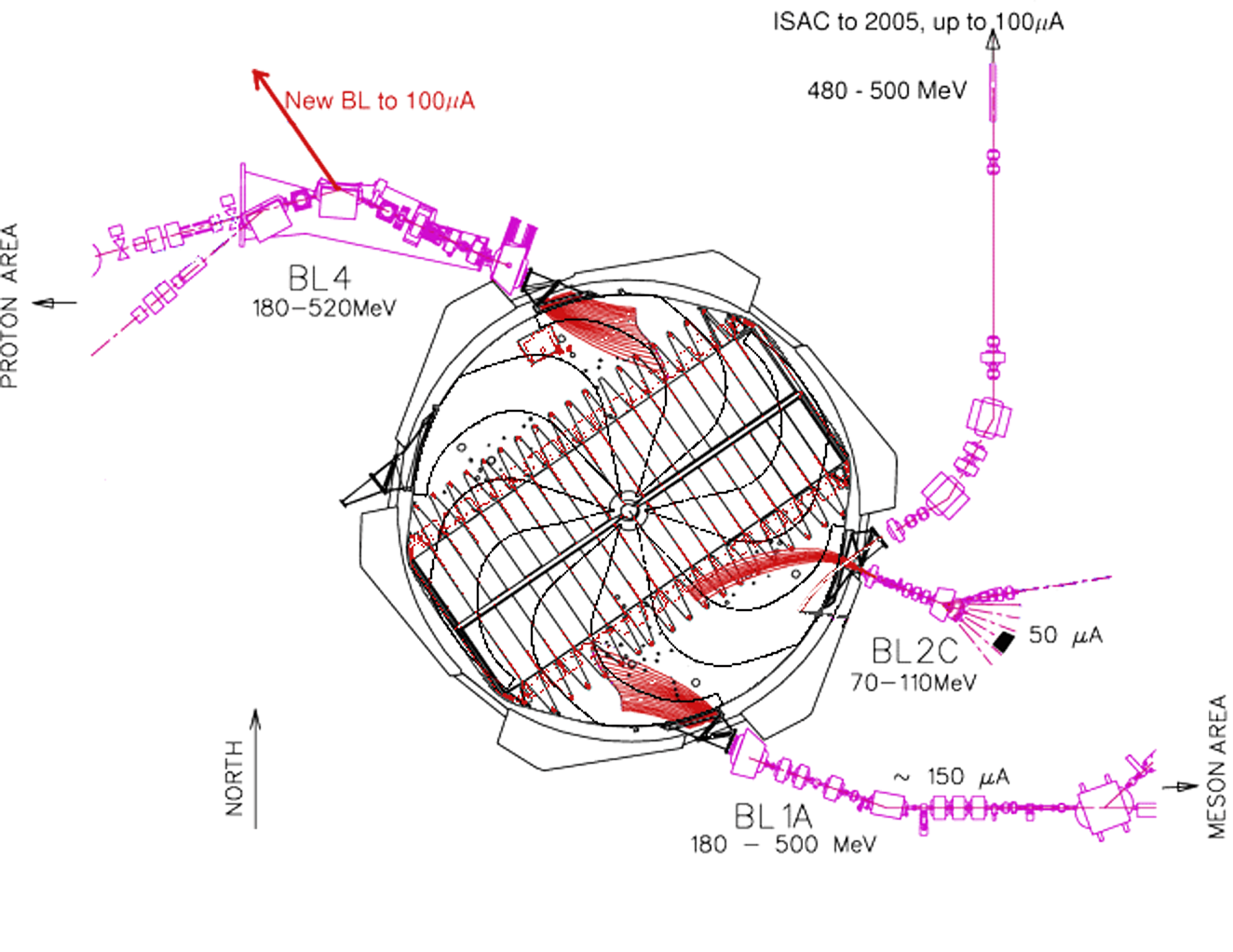}
\caption{Top view of the TRIUMF cyclotron, showing multiple
extraction paths of stripped protons (drawing provided by R.
Baartmann, TRIUMF, 2011)} \label{fig:triumf_extr}
\end{figure}

The other extraction scheme utilizes a stripping foil at the
extraction point. Accelerated ions change their charge state when
they pass through the foil. Owing to the change in the curvature
of the orbit, the beam can then be extracted from the cyclotron
bending field. This scheme can be applied to H$^-$ or H$_2^+$ ions
in order to produce a proton beam. However, the second electron of
the H$^-$ ion is weakly bound, and thus there exists a significant
probability that the electron will be detached from the ion in a
strong magnetic field. This effect generates unwanted losses from
the beam. H$^-$ ions are accelerated in the TRIUMF cyclotron; this
provides versatility in the form of options to extract the beam at
different energies, and even at several extraction points in
parallel (Fig.~\ref{fig:triumf_extr}). To limit the losses from
unwanted ionization, a moderate average field of 0.46~T was chosen
for this cyclotron. The H$_2^+$ ion has a stronger binding energy.
However, the bending field of the cyclotron must be stronger
because of the smaller charge-to-mass ratio of $1/2$
\cite{calabretta}.

In summary, charge stripping represents an elegant method of
extracting beams from a cyclotron. However, the following effects
can potentially limit the efficiency of the stripping method and
must be investigated.
\begin{itemize}
\item Scattering from residual gas molecules, and strong magnetic
fields can cause dissociation and thus loss of the accelerated
ions. \item The stripping process can also generate particles with
unwanted charge-to-mass ratios, for example neutrals, and the
transport and the locations of loss of these particles must be
considered. \item The lifetime of the stripping foil is typically
problematic; the foil is heated by power deposition from the beam
and also the stripped electrons, which are bound in the presence
of strong magnetic fields.
\end{itemize}

\section{Magnets}

\begin{figure}
\centering\includegraphics[width=0.50\textwidth]{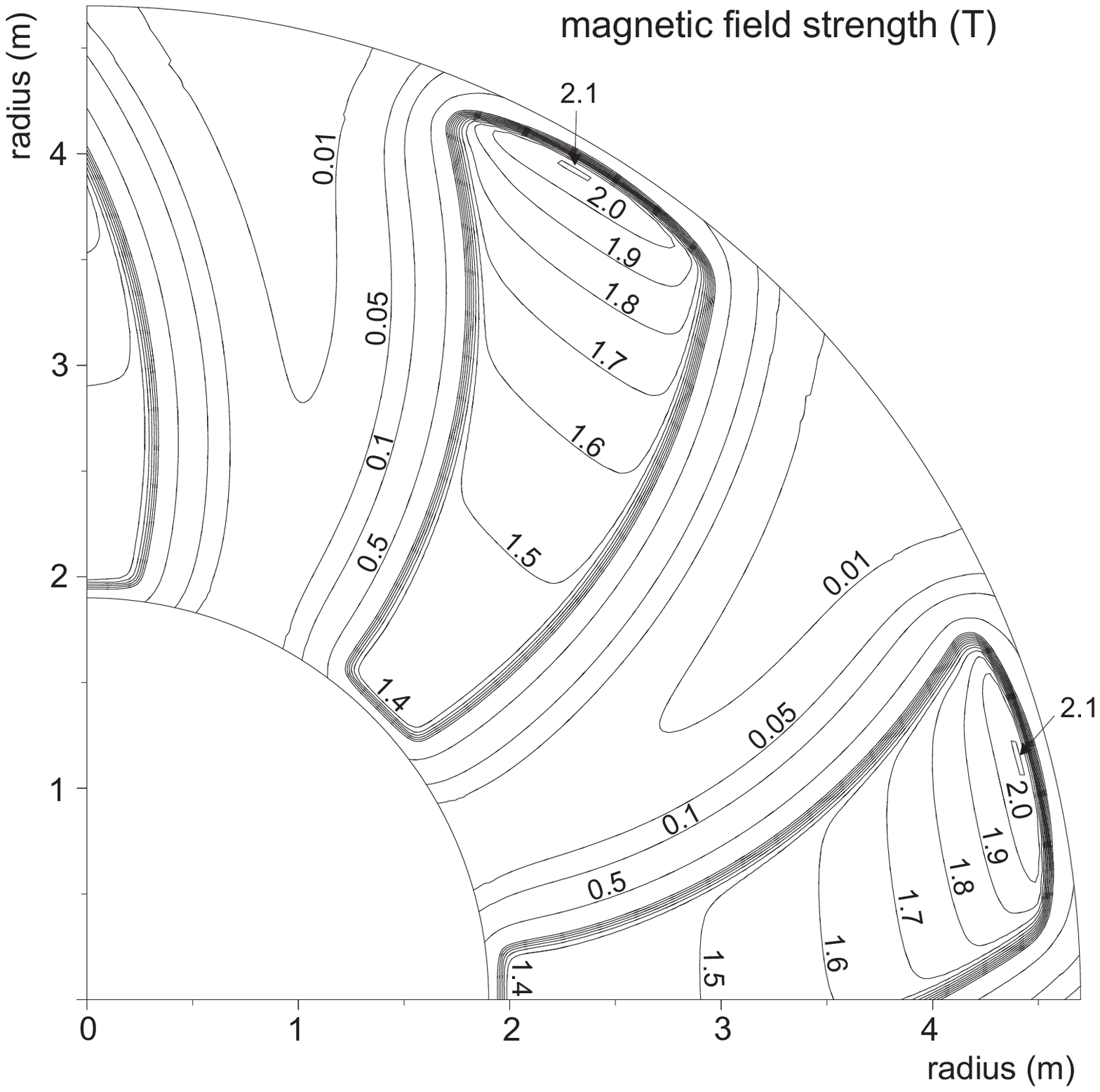}~
\includegraphics[width=0.49\textwidth]{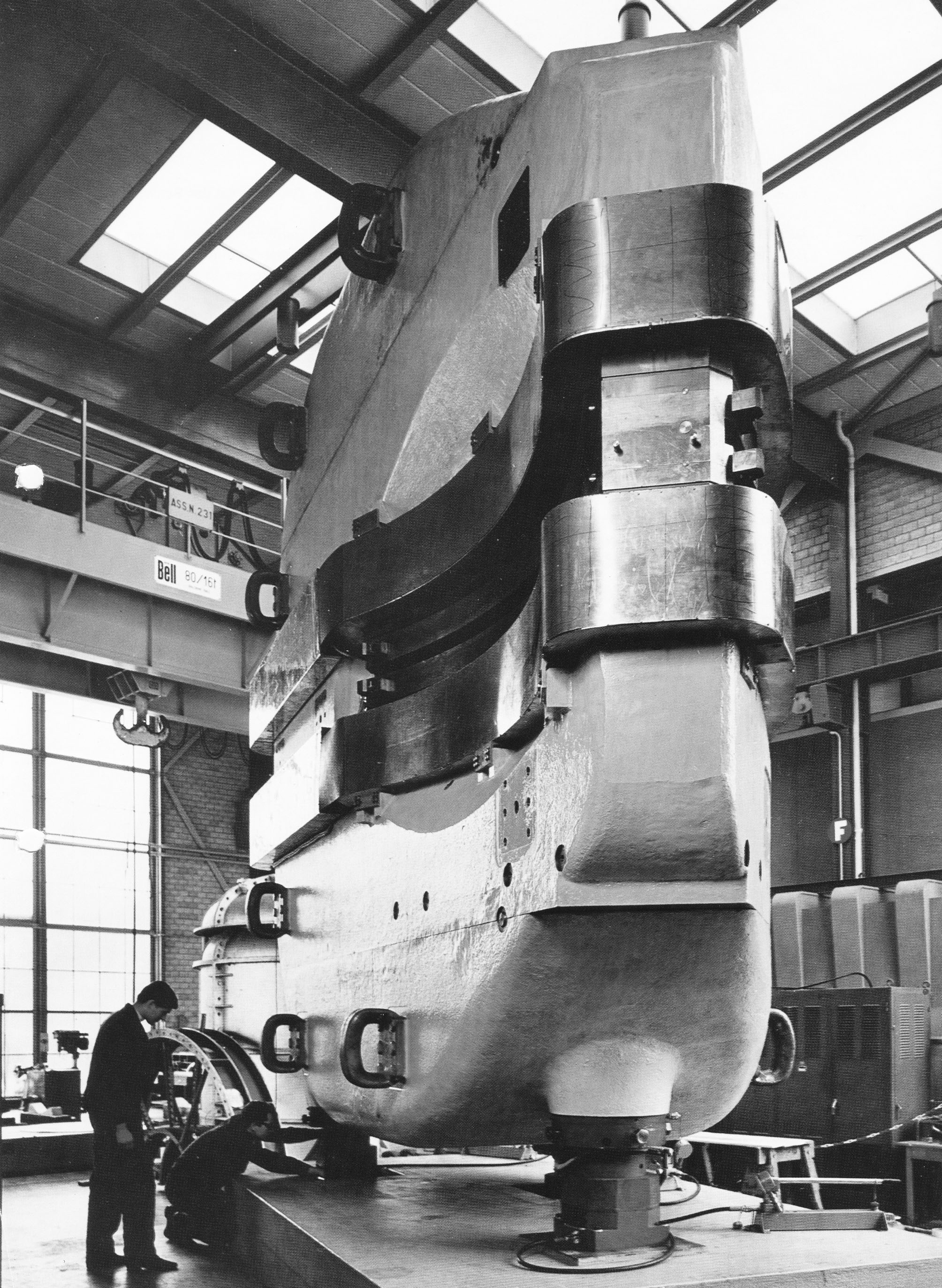} \caption{Left:
field map of the Ring cyclotron sector magnets at the PSI. The
field increases towards larger radii to keep the particle
revolution time constant. Right: photograph of a sector magnet
before installation. Note the curved shape of the pole edges.}
\label{fig:magnet} \end{figure}

As stated previously, the condition of isochronicity in a
cyclotron requires the average magnetic field to be increased in
proportion to $\gamma$ as the radius increases. The cyclotron
magnets have to be designed in such a way as to fulfil this
requirement. The increase in average field can be achieved by
introducing a slight vertical opening angle between the magnet
poles. For fine-tuning of the isochronicity, cyclotron magnets are
typically equipped with several correction coil circuits. Because
of the variation in the radius of the beam, which is typically
significant, cyclotron magnets have to cover a wide radial range.
The mechanical design becomes large and heavy. In the case of the
PSI Ring cyclotron, each magnet has a weight of 280~t.
 A field
map and a photograph of the sector magnets of the PSI Ring
cyclotron are shown as an example in Fig.~\ref{fig:magnet}.
Besides the purpose of bending the beam orbit, the magnets must
provide sufficient focusing in both planes. The magnets often have
a spiral shape for this purpose (see Section \ref{sec:avf}). Most
sector magnets for cyclotrons use normal-conducting coils,
although some medical and industrial cyclotrons use
superconducting magnets, to allow a compact and cost-effective
design. Joho~\cite{joho2} compared the weight statistics of
normal-conducting and superconducting cyclotrons. On average,
those using superconducting coils were lighter by a factor of 15.
The RIKEN cyclotron is unusual in this context, and employs
superconducting magnets with a peak field strength of 3.8~T
\cite{okuno}.

\section{Radio frequency systems}

\begin{figure}
\centering\includegraphics[width=0.5\textwidth]{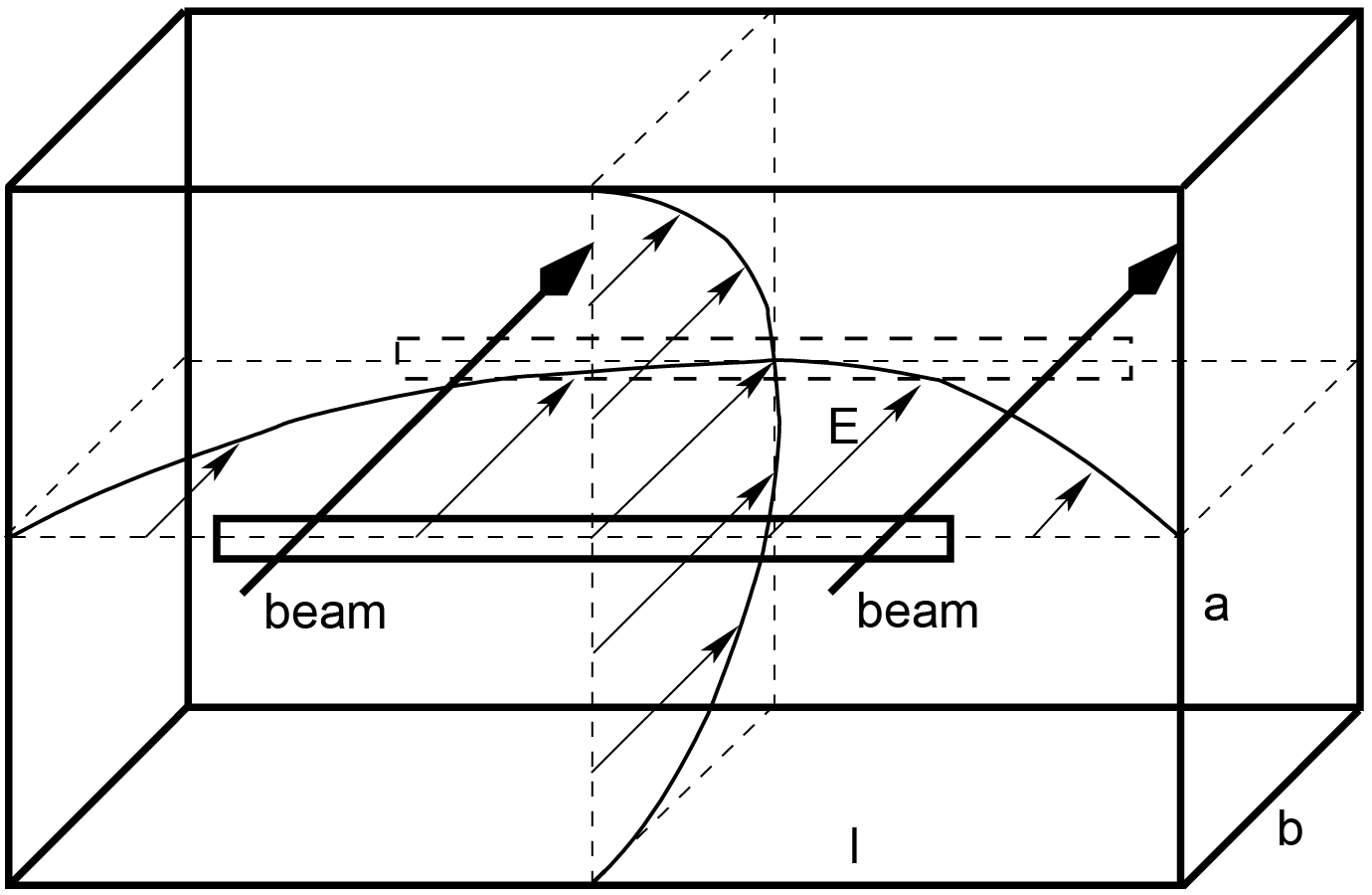}
\caption{Field distribution and orientation of the beam in a
cyclotron box resonator} \label{fig:boxcav} \end{figure}

In a classical cyclotron, an alternating voltage is applied across
the gap between the dees (Fig.~\ref{fig:spiral}). In a
separated-sector cyclotron, the space between the magnets allows
separate resonators to be installed. These resonators function in
principle like a rectangular cavity and can provide a
significantly higher gap voltage, for example 1~MV. The beam
passes through the resonator via a slit in the midplane. The
electric field strength varies as a sine function along the radius
(Fig.~\ref{fig:boxcav}).

\begin{figure}
\centering\includegraphics[width=0.92\textwidth]{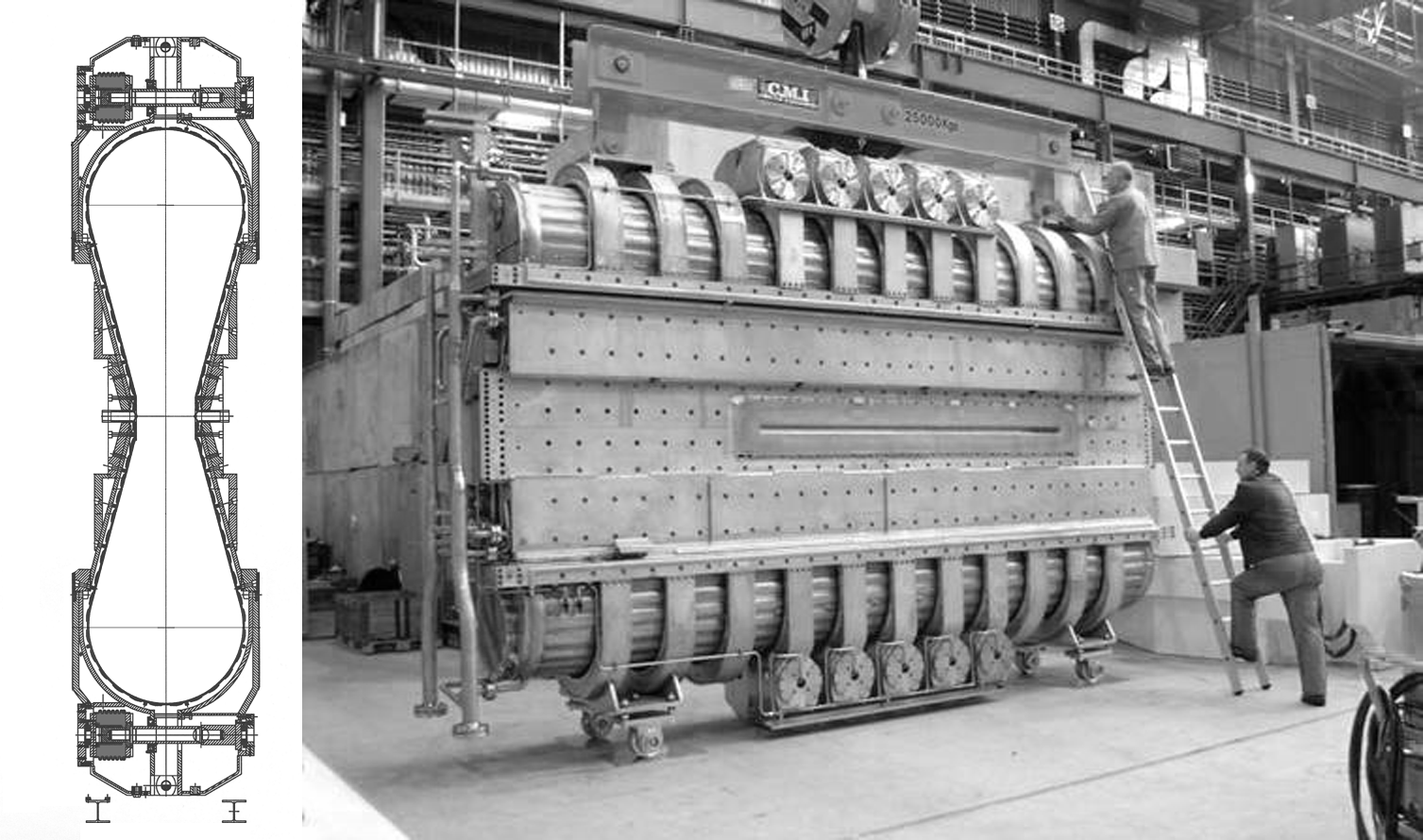}
\caption{Cross-section of PSI resonator (left), and photograph
(right)} \label{fig:cav_picture} \end{figure}

In this configuration, the resonance frequency depends on the
radial length $l$ and the height $a$ of the cavity:
\begin{equation}
f_0 = \frac{c}{2} \sqrt{\frac{1}{a^2} + \frac{1}{l^2}} .
\end{equation}
Thus the frequency is independent of the azimuthal width $b$ of
the cavity. In practice, the shape of the cavity is not made
exactly rectangular; instead, the azimuthal width is reduced in
the midplane (Fig.~\ref{fig:cav_picture}) to minimize the travel
time of the particles in the field. In the case of the PSI
cyclotrons, all accelerating resonators are operated at 50.6~MHz.
In the Ring cyclotron, the resonators are made from copper and
achieve a quality factor of $4.8\times 10^4$. At the time of
writing, the typical gap voltage is 830~kV. The design value is
higher, about 1.2~MV. Each resonator can transfer 400~kW of power
to the beam.

In order to minimize the variation of the energy over the bunch
length, cyclotrons are often equipped with so-called flat-top
resonators, which operate at the third harmonic. Using a
decelerating flat-top voltage at 1/9 of the amplitude of the
fundamental mode, the second derivative of the total voltage can
be made zero at the nominal phase. In this way, the variation of
the voltage as a function of the longitudinal position is
minimized. For high-intensity cyclotrons, the power transfer from
the electrical supply grid to the beam is a critical issue.
Typically, staged tube amplifiers are used to generate the
high-power RF signals required. The RF power is transferred to the
resonators via coaxial lines and is coupled to the resonator
volumes using loop couplers. The efficiency of the power transfer
can be estimated from the product of the individual efficiencies
of the components in the power transfer chain. The following
values have been determined for the PSI: AC/DC conversion, 0.90;
DC/RF conversion, 0.64; and RF to beam transfer, 0.55. Thus the
total efficiency is about 32\%, which is a relatively good number
for a particle accelerator.

\section{Performance of high-intensity cyclotrons, and discussion}

As discussed in the previous sections, the major limitation on the
operation of high-intensity cyclotrons is imposed by the
extraction losses. With a world record beam power of 1.3~MW
\cite{seidel} in the PSI Ring cyclotron, the relative losses are
at the level of $10^{-4}$. The majority of the lost protons hit
the vacuum chamber in the extraction beam line and activate
magnets and other accelerator components. After many years of
operation, the typical activation level is around 1~mSv/h, and
hotspots of approximately 10~mSv/h are observed in the extraction
beam line. In order to minimize the dose to personnel during
servicing of critical components, special mobile shielding devices
have been developed, for example for the exchange of the
electrostatic extraction channel. The ultimate criterion for the
activation problem in a high-intensity accelerator is the
radiation dose that the service personnel receive during
maintenance work. In the PSI facility, the total charge delivered
per year has significantly increased over the years. Nevertheless,
there exists no correlation with the dose received by personnel
\cite{tcads}. This fact demonstrates that it has been possible to
keep the absolute beam losses at a constant level.

Another important aspect of the performance of a high-intensity
accelerator concerns the efficiency of the power transfer from the
grid to the beam. The PSI accelerator complex consumes 10~MW in
total, and the beam power amounts to 1.3~MW. The total power
includes experimental facilities and many magnets that are not
essential for the production of the high-power beam. If one
considers only the RF systems, the overall efficiency is 32\%.
Remarkably, the majority of the beam power is transferred through
only four resonators in the Ring cyclotron. A potential
application of high-intensity proton accelerators is in
accelerator-driven systems (ADSs), which are subcritical reactors
for burning thorium \cite{rubbia} or the transmutation of nuclear
waste \cite{sheffield}. For such applications, the reliability and
trip rate (the rate of short beam interruptions) of the
accelerator are of the utmost importance. The typical trip rate of
the PSI accelerator in recent years has been in the range of
20--50 trips per day. Most trips are caused by electrical
breakdowns due to the voltage in electrostatic elements. After a
trip, the beam current is ramped back up to its nominal value
within 30~s. Trips of the RF systems occur much less frequently. A
statistical analysis of trip durations has been given in
Ref.~\cite{seidel2}. ADS systems require a much lower trip rate,
of the order of 0.01--0.1 trips per day. Although there exists a
promising potential for improvement (see also \cite{seidel2}), it
will be very difficult to achieve performance in this range
starting from today's performance.

The following particular advantages and disadvantages of the
cyclotron concept for high-intensity beam production can be
stated.
\begin{itemize}
\item A cyclotron requires an extraction element with an electrode
placed close to the beam. By comparison, an L-band superconducting
linac has a large aperture, and thus it is potentially easier to
achieve low losses in a linac. The electrostatic elements in
cyclotrons are critical and fragile devices, causing relatively
frequent beam trips and failures. \item Because of the concept of
a circular accelerator, the beam dynamics in a cyclotron is more
complicated, and it requires tedious tuning to achieve an
optimized operational state with low losses. \item For fundamental
reasons, the maximum energy of a cyclotron is limited to roughly
1~GeV. \item The edge focusing used in cyclotrons is weaker than
the alternating-gradient focusing in linacs. In particular, the
space charge forces in the vertical plane lead to a limitation on
the maximum beam current. It is expected that maximum currents in
the region of 10~mA can be achieved in sector cyclotrons
\cite{stammbach2}. \item On the pro side, the circular-cyclotron
concept allows the accelerating resonators to be re-used many
times. The footprint of a cyclotron facility is smaller, allowing
some savings with respect to the shielding and the building. \item
The low-frequency resonators in a cyclotron are robust, simple
devices with low trip rates and allow very high power throughput
in the couplers. \item The efficiency of the power transfer from
the grid to the beam is comparatively high, in the region of 30\%.
\end{itemize}

In summary, the cyclotron concept is capable of delivering
high-intensity beams with a beam power of up to 10~MW and an
energy of 1~GeV and represents an effective alternative to other
concepts in this range.


 \end{document}